# High-speed escape from a circular orbit

Philip R. Blanco[a)]
*Department of Physics and Astronomy, Grossmont College,
El Cajon, CA, 92020-1765
Department of Astronomy, San Diego State University,
San Diego, CA 92182-1221, USA*

Carl E. Mungan[b)]
*Physics Department, United States Naval Academy
Annapolis, MD, 21402-1363*

You have a rocket in a high circular orbit around a massive central body (a planet, or the Sun) and wish to escape with the fastest possible speed at infinity for a given amount of fuel. In 1929 Hermann Oberth showed that firing two separate impulses (one retrograde, one prograde) could be more effective than a direct transfer that expends all the fuel at once. This is due to the Oberth effect, whereby a small impulse applied at periapsis can produce a large change in the rocket's orbital mechanical energy, without violating energy conservation. In 1959 Theodore Edelbaum showed that this effect could be exploited further by using up to three separate impulses: prograde, retrograde, then prograde. The use of more than one impulse to escape can produce a final speed even faster than that of a fictional spacecraft that is unaffected by gravity. We compare the three escape strategies in terms of their final speeds attainable, and the time required to reach a given distance from the central body. To do so, in an Appendix we use conservation laws to derive a "radial Kepler equation" for hyperbolic trajectories, which provides a direct relationship between travel time and distance from the central body. The 3-impulse Edelbaum maneuver can be applied to interplanetary transfers, exploration of the outer solar system and beyond, and (in time reverse) efficient arrival and orbital capture. The physics principles employed are appropriate for an undergraduate mechanics course.

## I. INTRODUCTION

Newton's laws of motion and universal gravitation allow us to relate the geometrical properties of an orbit to conserved quantities such as angular momentum and energy. Changing the velocity of an orbiting object changes these quantities, and therefore the orbital path. One of the earliest descriptions of such orbital maneuvers was Newton's cannon,[1] a thought experiment that demonstrates the effects of changing a projectile's launch speed on its resulting orbit — increasing this speed increases the size and period of the orbit. For launch speeds greater than the local escape speed, the projectile does not return, but follows a hyperbolic escape path and approaches an asymptotic final speed $v_\infty$ far from the central body.

In the 20th century, pioneers of astrodynamics began to use these principles to chart the courses of hypothetical spacecraft that propel themselves by the directed expulsion of matter carried as fuel. The resulting impulses change their orbital paths. Even for a simple system consisting of a rocket and a massive central body, a variety of maneuvers are possible, such as Hohmann[2] and bi-elliptic transfers between orbits, with differing numbers of impulses employed to use the minimal amount of fuel.[3]

In this paper we discuss three strategies for a spacecraft in a circular orbit to achieve a high-speed escape with $v_\infty > 0$. We compare their fuel requirements to achieve a given $v_\infty$, and their travel times to reach a given distance from the central body.

In Sec. II we start with conservation laws to develop equations that relate orbital speed and distance. In Sec. III we use these to compare three escape strategies in terms of fuel usage. In Sec. IV, with the aid of an Appendix, we derive expressions for the travel time from the original orbit out to a given distance. In Sec. V we conclude by discussing applications and providing suggestions for student investigations.

## II. BASIC EQUATIONS FOR ORBITAL MANEUVERS

We assume that the mass of the spacecraft is tiny relative to the central body, such that any change in the latter's motion can be ignored. The central body can then be used to define an inertial frame in which we analyze the spacecraft's motions.

To simplify our analysis, we consider only impulses of duration much less than the orbital period around the central body. Equivalently, for such idealized *impulsive* maneuvers the rocket can change its velocity without changing its position, so that the pre- and post-impulse Keplerian orbits intersect. For such impulses to have the greatest effect on a spacecraft's orbital energy and angular momentum, they must be applied either prograde (in the direction of motion, by expelling exhaust backwards) or retrograde relative to the spacecraft velocity vector. This results in all paths being coplanar, reducing our analysis to two dimensions. (We discuss orbital plane changes briefly in Sec V.)

### 1. Speed as a function of distance

Between impulses, an object in freefall in the gravitational field of an isolated spherical body of mass $M$ will have a constant specific energy (where we use "specific" to mean "per unit mass"),

$$\varepsilon = \frac{1}{2}v^2 - \frac{GM}{r} \qquad (1)$$

Here, $v$ is the magnitude of its velocity vector, $r$ is its distance from the central body's center, and $G$ is the gravitational constant. (This energy properly belongs to the *system* of the spacecraft and central object, but since the massive central body is assumed stationary, we assign it to the spacecraft for brevity.)

From Eq. (1), in order to escape with an asymptotic final speed $v_\infty$ at infinity, an unpowered spacecraft at a distance $r_{\text{dep}}$ must have a departure speed $v_{\text{dep}}$ there given by

$$\varepsilon = \frac{1}{2}v_\infty^2 = \frac{1}{2}v_{\text{dep}}^2 - \frac{GM}{r_{\text{dep}}} \;\Rightarrow\; v_{\text{dep}} = \sqrt{\frac{2GM}{r_{\text{dep}}} + v_\infty^2}. \qquad (2)$$

For the maneuvers described in this paper, we also need to relate the orbital speed of a spacecraft to its position in a bound elliptical orbit. Let its speed be $v_A$ at *apoapsis* (farthest from the central body, at distance $r_A$) and $v_P$ at *periapsis* (the closest point, distance $r_P$). Due to the central nature of the gravitational field, the specific angular momentum is conserved and given by

$$h = r v_\theta, \qquad (3)$$

where $v_\theta$ is the azimuthal component of the spacecraft's velocity. Therefore, at the extremes of the ellipse where there is no radial motion,

$$h = r_A v_A = r_P v_P \; r v_\theta. \qquad (4)$$

Combining Eqs. (1) and (4) we can solve for apoapsis and periapsis speeds in terms of the corresponding distances

$$v_A = \sqrt{\frac{GM}{r_A}}\sqrt{\frac{2}{1+\frac{r_A}{r_P}}} \text{ and } v_P = \sqrt{\frac{GM}{r_P}}\sqrt{\frac{2}{1+\frac{r_P}{r_A}}}. \qquad (5)$$

The first square-root factor in each equation is the circular

### 2. Relating impulse to fuel usage and energy input

Typically, a rocket engine is modeled as one that expels its fuel continuously, even for short-duration impulses that represent the idealized limit of minimizing fuel usage and transfer times. Conservation of momentum leads to the *rocket equation*, attributed to Konstantin Tsiolkovsky, which relates the speed change $\Delta v$ to the initial and remaining mass of the rocket,

$$\Delta v = v_{\text{ex}} \ln \frac{m_i}{m_f}, \qquad (6)$$

where $m_i$ and $m_f$ are the masses of the rocket (including fuel) before and after the impulse, respectively, and $v_{\text{ex}}$ is the effective exhaust speed.[4] (The gravitational attraction between the rocket and exhaust mass can be ignored.) Equation (6) does not depend on the rate of fuel consumption, and applies to multiple intermittent firings of the rocket engine. For this reason, we shall follow the standard convention of using total $\Delta v$ as a proxy for fuel cost when comparing strategies to achieve an orbital transfer.

When a mass of fuel $(m_i - m_f)$ is expelled to obtain a given $\Delta v$, the chemical energy converted to kinetic energy is $\Delta E_{\text{chem}} = \frac{1}{2}(m_i - m_f)v_{\text{ex}}^2$. A fraction of this input energy changes the mechanical energy of the rocket, while the rest is carried away in the exhaust,[4] as we shall discuss in Sec. III.5.

## III. ESCAPE FROM A CIRCULAR ORBIT

Initially, we have a spacecraft in a circular orbit of radius $r_0$ with orbital speed $v_0 = \sqrt{GM/r_0}$. (This relation can be proven by centripetal arguments, or by setting $r_A = r_P = r_0$ in Eq. 5.) We desire to achieve a target final speed $v_\infty$ for the smallest total $\Delta v$, so that the smallest amount of fuel is required.

### 1. Single impulse for direct escape

The simplest escape strategy is for the rocket to fire a single, prograde impulse from the circular orbit, to increase its speed past the local escape speed $\sqrt{2}v_0$, as shown in Fig. 1(a). To find the specific impulse $\Delta v_D$ for such a direct escape to produce a final speed $v_\infty$, use Eq. (2) with $v_{\text{dep}} = v_0 + \Delta v_D$, and $r_{\text{dep}} = r_0$, i.e.

$$v_0 + \Delta v_D = \sqrt{\frac{2GM}{r_0} + v_\infty^2} \Rightarrow \Delta v_D = v_0\left(\sqrt{2 + \frac{v_\infty^2}{v_0^2}} - 1\right). \quad(7)$$

This function is plotted as a green dotted line in Fig. 2.

### 2. Two-impulse escape - the "Oberth maneuver"

In 1929, Hermann Oberth described a more fuel-efficient strategy[5] to obtain the same $v_\infty$ from a high circular orbit (of radius many times that of the central body). First use a retrograde impulse to drop closer to the body, then apply a prograde impulse at the periapsis of that elliptical orbit, as shown in Fig. 1(b). Author Robert Heinlein (who had worked with Oberth on the pioneering science fiction film *Destination Moon*) described this maneuver in chapter 7 of his novel *The Rolling Stones*:

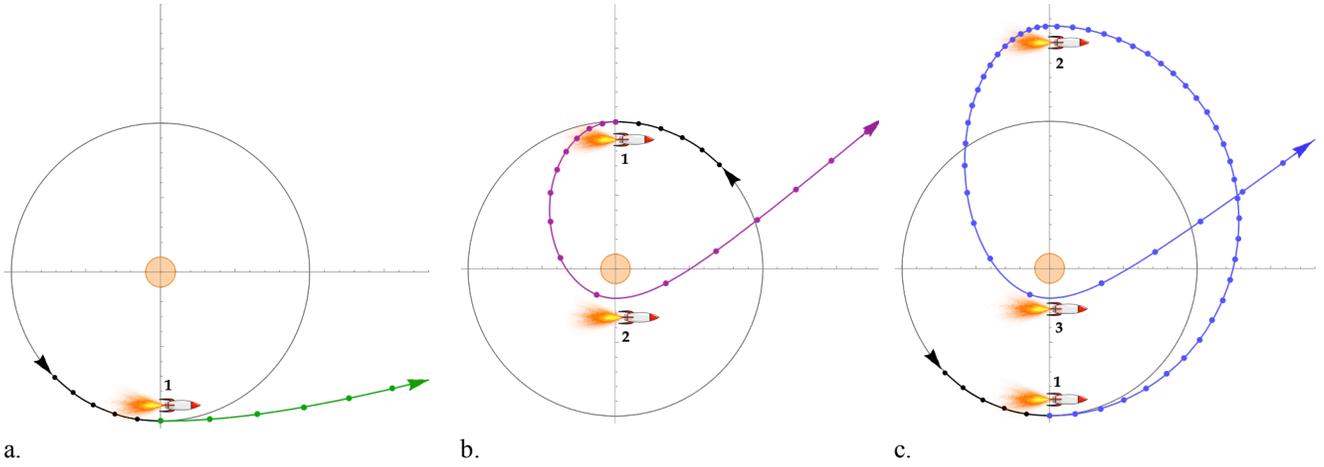

Fig. 1. Three escape strategies from a counter-clockwise circular orbit, all with a total $\Delta v = 1.1\ v_0$. Dots show the spacecraft position at intervals of 1/40th of the original circular orbital period. The rocket images mark the position and direction of each impulsive thrust. (a) Direct (single-impulse) escape. (b). Oberth (2-impulse) escape with $r_{in} = 0.15\ r_0$. (c) Edelbaum (3-impulse) escape with $r_{out} = 1.50\ r_0$ and $r_{in} = 0.15\ r_0$. (These values were chosen for clarity of plotting, and are not the same as those adopted for Figs. 2 and 3.)

"A ship leaving the Moon or a space station for some distant planet can go faster on less fuel by dropping first toward the Earth, then performing her principal acceleration while as close to the Earth as possible." [6]

Reference 7 explains the physics of this *Oberth effect* in detail. If a rocket body of mass $m$ moves at speed $v$, a forward specific impulse $\Delta v$ increases its kinetic energy by $\frac{1}{2}m(v + \Delta v)^2 - \frac{1}{2}mv^2$, which increases linearly with $v$. Since an orbiting rocket travels faster as it moves deeper into the potential well, firing an impulse as close as possible to the central body allows it to maximize the energy gained, in principle without limit as $r_P \to 0$. (In practice $r_P$ is limited by the size of the central body).

The increase in the mechanical energy of the *system* of rocket+exhaust comes from the chemical energy converted in the impulse $\Delta E_{chem}$, which is constant regardless of where the impulse was fired. Therefore, any additional energy gained by a fast-moving rocket must be obtained at the expense of the mechanical energy carried by the expelled fuel, as we discuss in Sec. III.5.

The two-impulse Oberth maneuver is defined by the periapsis distance $r_{in}$ of the elliptical orbit produced by the initial, retrograde impulse at radius $r_0$. From Eq. (5) with $r_A = r_0$ and $r_P = r_{in}$, the required velocity change for this first impulse $\Delta v_{Ob1}$ is

$$\Delta v_{Ob1} = v_A - v_0 = -v_0\left(1 - \sqrt{\frac{2}{1+\frac{r_0}{r_{in}}}}\right), \quad (8)$$

where the leading negative sign denotes the retrograde direction. Once the spacecraft reaches periapsis, it will be moving at speed $v_P$ given by Eq. (5) with $r_A = r_0$ and $r_P = r_{in}$. To achieve a final speed $v_\infty$, use Eqs. (2) and (5) with $r_{dep} = r_{in}$ to find the second prograde velocity boost $\Delta v_{Ob2}$ from $v_P$ to $v_{dep}$,

$$\Delta v_{Ob2} = v_{dep} - v_P = \sqrt{\frac{GM}{r_{in}}}\left(\sqrt{2 + \frac{r_{in}v_\infty^2}{GM}} - \sqrt{\frac{2}{1+\frac{r_{in}}{r_0}}}\right). \quad (9)$$

The total specific impulse for the Oberth escape maneuver is the sum of the absolute values of Eqs. (8) and (9), i.e. $\Delta v_{Ob} = |\Delta v_{Ob1}| + \Delta v_{Ob2}$, which simplifies to

$$\Delta v_{Ob} = v_0\left(1 + \sqrt{\frac{v_\infty^2}{v_0^2} + \frac{2r_0}{r_{in}}} - \sqrt{2 + \frac{2r_0}{r_{in}}}\right). \quad (10)$$

Equation (10) is plotted in Fig. 2 as a dashed purple curve for $r_{in} = 0.05r_0$. This curve crosses that described by Eq. (7) at $\Delta v_{Ob} = \Delta v_D = v_0$, independent of $r_{in}$, for which the resulting $v_\infty = \sqrt{2}v_0$, coincidentally equal to the local escape speed from the original orbit. For desired values of $v_\infty$ larger than this, $\Delta v_{Ob} < \Delta v_D$, so the two-impulse Oberth maneuver will require less fuel than the single-impulse direct escape (cf. Fig. 2).

As $r_{in}$ is decreased, the "Oberth advantage" (the difference $\Delta v_D - \Delta v_{Ob}$) gets larger for $v_\infty > \sqrt{2}v_0$, since the fuel expelled at periapsis is placed into a lower energy orbit. In the theoretical limit as $r_{in} \to 0$, the dashed purple curve in Fig. 2 becomes flat, and Eq. (8) shows that $\Delta v_{Ob1} \to -v_0$ for any $v_\infty$. This speed change stops the spacecraft in its original orbit and causes it to fall in radially towards the center of attraction. Then, the tiniest boost $\Delta v_{Ob2}$ at $r_{in} = 0$ could produce any value of $v_\infty$ desired.

### 3. Three-impulse escape - the "Edelbaum maneuver"

In 1959 Theodore Edelbaum described a coplanar escape maneuver that employs three impulses: a prograde boost to raise apoapsis, followed by a retrograde impulse to cause the spacecraft to fall in to a low periapsis, and thence a final boost to escape. [8] An example is shown in Fig. 1(c). By falling in from a larger radius than the original circular orbit, the resulting faster periapsis speed enhances the Oberth effect, by transferring more of the expelled fuel's chemical *and* mechanical energy to the spacecraft[7].

The transfer from circular orbit to the intermediate ellipse is defined by the apoapsis radius of that ellipse, $r_A = r_{out}$, with periapsis at the original orbit radius, $r_P = r_0$. Equation (5) gives the required post-impulse velocity $v_P$, from which we find the necessary boost $\Delta v_{Ed1}$,

$$v_P = v_0 + \Delta v_{Ed1} \Rightarrow \Delta v_{Ed1} = v_0\left(\sqrt{\frac{2}{1+\frac{r_0}{r_{out}}}} - 1\right). \quad (11)$$

The resulting apoapsis speed is also found from Eq. (5). Once the spacecraft reaches this position, it fires a second, retrograde impulse to lower its periapsis distance to a new $r_P = r_{in} < r_0$. The required velocity change $\Delta v_{Ed2}$ is then

$$\Delta v_{Ed2} = -\sqrt{\frac{GM}{r_{out}}}\left(\sqrt{\frac{2}{1+\frac{r_{out}}{r_0}}} - \sqrt{\frac{2}{1+\frac{r_{out}}{r_{in}}}}\right). \quad (12)$$

The speed $v_P$ of the rocket after it falls from apoapsis at $r_{out}$ to periapsis at $r_{in}$ is given by Eq. (5), and the required velocity for escape from there to a specified $v_\infty$ is found from Eq. (2) with $r_{dep} = r_{in}$. The difference is the final velocity boost $\Delta v_{Ed3}$, i.e.

$$\Delta v_{Ed3} = v_{dep} - v_P = \sqrt{\frac{GM}{r_{in}}}\left(\sqrt{2 + \frac{r_{in}v_\infty^2}{GM}} - \sqrt{\frac{2}{1+\frac{r_{in}}{r_{out}}}}\right). \quad (13)$$

The total speed change for an Edelbaum escape is the sum of the absolute values of Eqs. (11), (12), and (13), $\Delta v_{Ed} = \Delta v_{Ed1} + |\Delta v_{Ed2}| + \Delta v_{Ed3}$, and can be simplified to

$$\Delta v_{Ed} = v_0\left(\sqrt{\frac{v_\infty^2}{v_0^2} + \frac{2r_0}{r_{in}}} + \sqrt{2 + \frac{2r_0}{r_{out}}} - \sqrt{\frac{2r_0}{r_{in}} + \frac{2r_0}{r_{out}}} - 1\right). \quad (14)$$

Equation (14) is plotted in Fig. 2 as a blue dot-dashed curve for $r_{in} = 0.05r_0$ and $r_{out} = 2.5r_0$. It is also valid for computing $\Delta v_{Ob}$ if $r_{out} = r_0$, and for $\Delta v_D$ if $r_{out} = r_{in} = r_0$.

Compared to the two-impulse Oberth escape maneuver, the Edelbaum escape always has a lower overall $\Delta v$ for a given $v_\infty$ by an amount
$\Delta v_{Ob} - \Delta v_{Ed} =$

$$\sqrt{2}v_0\left(\sqrt{2} + \sqrt{\frac{r_0}{r_{in}} + \frac{r_0}{r_{out}}} - \sqrt{1 + \frac{r_0}{r_{out}}} - \sqrt{1 + \frac{r_0}{r_{in}}}\right), \quad (15)$$

independent of $v_\infty$, with a corresponding savings in fuel expenditure. (Readers can show that Eq. 15 is positive for any $r_{in} < r_0$ and $r_{out} > r_0$.) This difference can be seen as the constant vertical distance between the dashed purple (Oberth) and dot-dashed blue (Edelbaum) curves in Fig. 2.

In Fig. 2, the dot-dashed blue curve for $\Delta v_{Ed}$ crosses the dotted green curve for $\Delta v_D$ when
$\Delta v_{Ed} = \Delta v_D$

$$= v_0\left(\sqrt{2 + \frac{2r_0}{r_{out}}} - 1\right) \text{ for } v_\infty = \sqrt{\frac{2r_0}{r_{out}}}v_0. \quad (16)$$

For larger desired values of $v_\infty$, the Edelbaum maneuver requires a lower total $\Delta v$ than the direct or Oberth escapes, and so is fuel-optimal for such high-speed escapes.

In the dual limit $r_{out} \to \infty$ and $r_{in} \to 0$, $\Delta v_{Ed} \to (\sqrt{2} - 1)v_0$. This corresponds to a transfer which first sends the spacecraft very far from the central body, then (much later) applies a very small retrograde impulse to nullify its angular momentum, causing it to fall radially inwards to $r_{in} = 0$, where a tiny impulse can produce any value of $v_\infty$ desired. While this *bi-parabolic* trajectory has the smallest possible $\Delta v$ ($\approx 0.41v_0$) to achieve any desired $v_\infty$, it requires an infinite travel time.[9,10]

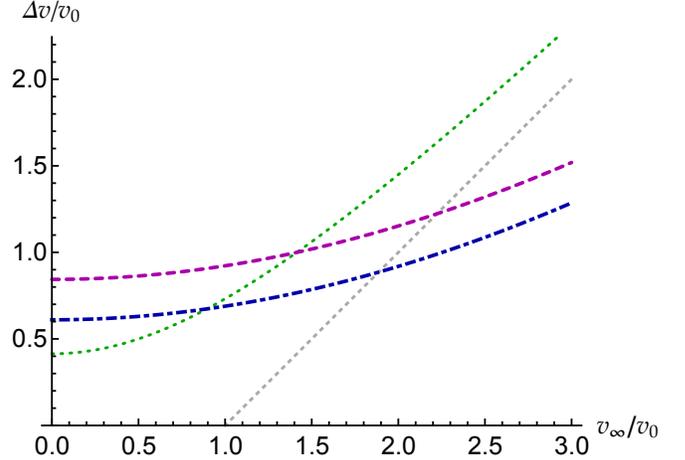

Fig. 2. Total required speed change $\Delta v$ for the three escape strategies as a function of desired final speed $v_\infty$, both normalized by the original circular speed $v_0$. Dotted green curve: Direct escape from Eq. (7). Dashed purple curve: Oberth (2-impulse) escape with $r_{in} = 0.05 r_0$ from Eq. (10). Dot-dashed blue curve: Edelbaum (3-impulse) escape with $r_{out} = 2.5 r_0$ and $r_{in} = 0.05 r_0$ from Eq. (14). Dotted gray line: "no gravity" escape $\Delta v = v_\infty - v_0$ from Sec. III.6.

We can invert and simplify Eq. (14) to give an expression for the mission's final $v_\infty$ for an available overall $\Delta v$ provided by the fuel,
$v_\infty^2 =$

$$v_0^2\left[\left(1 + \frac{\Delta v}{v_0} + \sqrt{\frac{2r_0}{r_{in}} + \frac{2r_0}{r_{out}}} - \sqrt{2 + \frac{2r_0}{r_{out}}}\right)^2 - \frac{2r_0}{r_{in}}\right]. \quad (17)$$

This relation also applies to the 2-impulse Oberth escape if one sets $r_{out} = r_0$, and to the direct escape if one sets $r_{out} = r_{in} = r_0$.

### 4. Four or more impulses?

A valid question is whether additional fuel savings can be achieved using more than three impulses. It has been shown that any *fuel-optimal* coplanar transfer will consist of no more than three impulses.[10] This also holds true for coplanar bound orbit-to-orbit transfers, which are either Hohmann (two impulses)[2] or bi-elliptic (three).[3] Coplanar escape strategies that produce fuel-optimal transfers with more than three impulses can always be reduced to an equivalent with three or fewer.[11,12]

As an example, consider the following 4-impulse strategy: (1) first slow down from circular orbit to an intermediate inner radius $r_{in}$, (2) use the Oberth effect there to boost to $r_{out}$, (3) slow down again to lower periapsis to $r_{in}$, prior to (4) boosting to escape. We have shown above that more efficient escapes are realized as $r_{out} \to \infty$ and $r_{in} \to 0$ (to increase periapsis speed and hence the Oberth advantage of the final impulse). Therefore, any intermediate ellipse can be made more fuel-efficient by increasing $r_{out}$ and decreasing $r_{in}$ to their mission-constrained extremes. After doing so, the optimal 4-impulse escape degenerates to the same path as the 3-impulse escape and is no more efficient.

For idealized impulsive maneuvers, a rocket in a closed orbit will always return to the position at which the impulse was applied (unless another impulse is made). A prograde or retrograde impulse at one apsis of an orbit raises or lowers the other apsis. It further follows from Eq. (6) that multiple impulses fired at the same position in orbit are fuel-equivalent to a single, combined impulse.

For example, if the rocket engine cannot apply the retrograde impulse for the Oberth or Edelbaum maneuver all at once, it can lower its periapsis successively to $r_{in}$ by multiple firings each time it passes through apoapsis, without any loss of overall fuel efficiency. Students can use the equations in Sec. II to design and evaluate their own multi-impulse maneuvers.

### 5. Energetics of impulsive escape maneuvers

How can the same fuel expenditure (and hence energy converted $\Delta E_{chem}$) produce different final rocket speeds $v_\infty$ for the three escape strategies? The initial mechanical energy of the rocket+fuel system of mass $m_i$ in a circular orbit is $E_{sys,i} = -\frac{1}{2}m_i v_0^2$ from Eq. (1). After all the fuel is expelled, the mechanical energy added to the system is $\Delta E_{chem}$ from Sec. II.2, which provides the final rocket mass $m_f$ with a speed change $\Delta v$ given by Eq. (6). The final mechanical energy $E_{sys,f}$ of the system of rocket+exhaust is then

$$E_{sys,f} = E_{sys,i} + \Delta E_{chem} = -\frac{1}{2}m_i v_0^2 + \frac{1}{2}(m_i - m_f)v_{ex}^2$$

$$= -\frac{1}{2}m_i v_0^2 + \frac{1}{2}m_i v_{ex}^2 \left[1 - \exp\left(-\frac{\Delta v}{v_{ex}}\right)\right], \quad (18)$$

which is constant for a given $\Delta v$ and fuel mass expended, and independent of the number of impulses. Therefore, if one wants the rocket to end up with the highest possible mechanical energy $m_f \varepsilon = \frac{1}{2}m_f v_\infty^2$, one should choose an impulse strategy that leaves the combined exhaust masses with the lowest.

As an exercise, by expressing $v_{ex}$ as a multiple of $v_0$, students can calculate $\Delta E_{chem}$ and the change in kinetic energy of the rocket (due to changes in its speed *and* mass) after each impulse. The difference gives the mechanical energy change (positive, negative, or zero) of the fuel mass expelled. This will reveal for each strategy how the same final system energy $E_{sys,f}$ is distributed between the escaping rocket and the total mass of expelled fuel.

### 6. Comparison with a "no gravity" rocket

A remarkable result is that both Oberth and Edelbaum escapes can attain a higher $v_\infty$ than a fictional rocket that is unaffected by the central body's gravity, which boosts tangentially from orbit at $r_0$ to a constant speed $v_\infty = v_0 + \Delta v$, shown as a dotted gray line in Fig. 2. For example, use Eq. (17) to calculate $v_\infty$ for each escape strategy using a mission total $\Delta v = 1.25 v_0$. With $r_{in} = 0.05 r_0$, the Oberth escape gives $v_\infty = 2.30 v_0$; additionally setting $r_{out} = 2.5 r_0$ gives $v_\infty = 2.92 v_0$ for the Edelbaum escape — both faster than $v_\infty = 2.25 v_0$ for the "no gravity" rocket.

The Oberth and Edelbaum trajectories can end up with larger values of $v_\infty$ because the fuel on board the "no gravity" rocket carries no gravitational potential energy relative to the central body. With no potential energy to "steal" from the expelled fuel, that spacecraft's kinetic energy gain is limited to a fraction of the fuel's chemical and kinetic energy only.[7]

## IV. TRAVEL TIME COMPARISON

Compared to a single-impulse direct escape, both the Oberth and Edelbaum escapes can produce faster final speeds for a given total $\Delta v$ (and thus fuel). However, these maneuvers require additional elliptical orbit segments where the spacecraft moves slowly. For very distant destinations ($r \gg r_0$), the extra time on these segments may not be important, but for intermediate distances there will be a trade-off in total time to destination. In this section we calculate the travel times from the initial circular orbit to a specified distance $r$ from the central body, to determine which of the three strategies will reach that distance in the shortest time.

### 1. Hyperbolic segment

All three escape strategies culminate in a hyperbolic escape from a prograde impulse applied at periapsis. For the direct escape, the departure radius $r_{dep} = r_0$; for the Oberth and Edelbaum escapes, $r_{dep} = r_{in}$.

In the Appendix, we derive a general relationship for a hyperbolic orbit between the time since periapsis and the distance $r$ from the central body. Equation (A6) gives the time of flight from a periapsis distance $r_P$ out to a destination distance $r$. We can rewrite that equation in terms of the original circular speed $v_0$ and period $T_0 = 2\pi r_0/v_0$,

$$t_{hyp}(r_P, r) = \frac{T_0}{2\pi} \frac{v_0^3}{v_\infty^3} \left[ \sqrt{\left(1 + \frac{r v_\infty^2}{r_0 v_\infty^2}\right)^2 - \left(1 + \frac{r_P v_\infty^2}{r_0 v_\infty^2}\right)^2} - \cosh^{-1}\left(\frac{1 + \frac{r v_\infty^2}{r_0 v_\infty^2}}{1 + \frac{r_P v_\infty^2}{r_0 v_\infty^2}}\right) \right]. \quad (19)$$

Table 1. Travel times for the escape strategies discussed in this paper, starting from a circular orbit of radius $r_0$ and ending at a final distance $r$ from the central body.

| # Impulses | $\Delta v$ | Total travel time (Eqs. 19-21) |
|---|---|---|
| 1 (direct) | Eq. (7) | $t_{\text{hyp}}(r_0, r)$ |
| 2 (Oberth) | Eq. (10) | $t_{\text{ell}}(r_0, r_{\text{in}}) + t_{\text{hyp}}(r_{\text{in}}, r)$ |
| 3 (Edelbaum) | Eq. (14) | $t_{\text{ell}}(r_0, r_{\text{out}}) + t_{\text{ell}}(r_{\text{out}}, r_{\text{in}}) + t_{\text{hyp}}(r_{\text{in}}, r)$ |
| 1 (no gravity) | $v_\infty - v_0$ | $t_{\text{nog}}(r_0, r)$ |

### 2. Elliptical segments (2- and 3-impulse escapes)

For the direct impulse, there are no intermediate orbits so the total travel time is just the hyperbolic segment time $t_{\text{hyp}}(r_0,r)$. For the Oberth and Edelbaum transfers, Kepler's third law provides the time spent on the elliptical segments. The transfer time $t_{\text{ell}}(r_A, r_P)$ between apoapsis and periapsis is half the period of an orbit of semi-major axis $(r_A + r_P)/2$, as seen in Fig. 1,

$$t_{\text{ell}}(r_A, r_P) = \frac{1}{2}\sqrt{\frac{4\pi^2}{GM}\left(\frac{r_A+r_P}{2}\right)^3} = \frac{T_0}{2}\sqrt{\left(\frac{r_A+r_P}{2r_0}\right)^3}. \quad (20)$$

To obtain the total travel time for the two-impulse Oberth maneuver, we must include the time for the infalling segment $t_{\text{ell}}(r_0, r_{\text{in}})$. For the three-impulse Edelbaum maneuver, we must include two elliptical segment transfer times $t_{\text{ell}}(r_0, r_{\text{out}}) + t_{\text{ell}}(r_{\text{out}}, r_{\text{in}})$.

### 3. Shortest travel time for Edelbaum escape

Expressions for the total travel times for the three escape strategies are summarized in Table I. These are plotted in Fig. 3 as a function of the Edelbaum maneuver "swing out" radius $r_{\text{out}}$, using the same parameters as for Fig. 2 and a destination distance $r = 200r_0$. Clearly for this case there is an optimal swing-out radius $r_{\text{out}} \approx 2.5r_0$ for the Edelbaum transfer to minimize the travel time. For larger values of $r_{\text{out}}$, the additional travel time in the first elliptical segment of Fig. 1(c) offsets the Oberth effect's advantage of gaining a larger $v_\infty$ for the same overall $\Delta v$.

There is no tractable closed-form solution for the time-minimizing $r_{\text{out}}$ as a function of destination distance $r$ and mission constraints $\Delta v_{\text{Ed}}$ and $r_{\text{in}}$. Instead, students can use Eq. (17), followed by Eqs. (19) and (20), to evaluate travel times for a range of values of $r_{\text{out}}$ given the values of the other parameters. As the destination distance $r$ increases, the time-optimal value of $r_{\text{out}}$ also increases, since the elliptical segments take up a smaller fraction of the overall trip duration, while the larger $v_\infty$ reduces the remaining time spent on the hyperbolic segment.

### 4. Travel times compared

Despite their slow starts, spacecraft executing Oberth and Edelbaum escapes can overtake a spacecraft on a direct escape and arrive at a distant destination sooner, as shown in Fig. 3. As discussed in Sec. III.6, a spacecraft on either of the 2- or 3-impulse trajectories can even overtake a fictional rocket that experiences no gravitational attraction to the central body. The travel time $t_{\text{nog}}$ for this putative "no gravity" rocket is simply the length of the linear segment from $r_0$ to $r$ divided by its constant $v_\infty = v_0 + \Delta v$,

$$t_{\text{nog}}(r_0, r) = \frac{\sqrt{r^2 - r_0^2}}{v_\infty} = \frac{T_0}{2\pi}\frac{\sqrt{\frac{r^2}{r_0^2}-1}}{1+\frac{\Delta v}{v_0}}, \quad (21)$$

and is shown as the dotted gray line in Fig. 3.

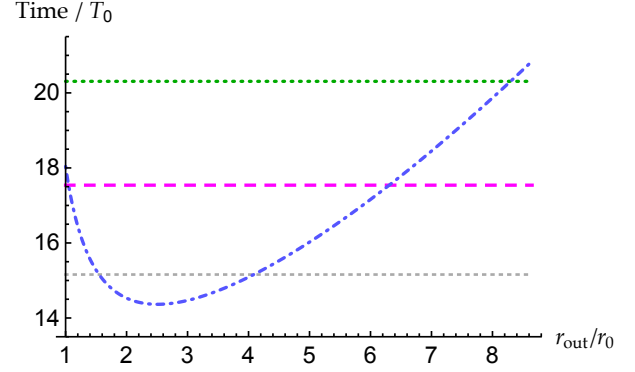

Fig. 3. Time to reach $r = 200\, r_0$ (as a multiple of the original circular orbit period $T_0$) for the 3-impulse Edelbaum escape (dot-dashed blue curve) as a function of its normalized swing-out radius $r_{\text{out}}/r_0$, for $\Delta v = 1.1\, v_0$ and $r_{\text{in}} = 0.05\, r_0$. Travel times using these values are also shown for the direct (dotted green line), Oberth (dashed purple line), and "no gravity" (dotted gray line) escapes, computed according to Table 1.

## V. DISCUSSION

For a spacecraft that can barely escape the central body ($v_\infty \approx 0$), Fig. 2 shows that a single-impulse direct transfer is most efficient in terms of $\Delta v$. This is also the case for a spacecraft in a low orbit that cannot approach the central body any closer than its original orbital radius $r_0$. For a high-speed escape from a high circular orbit, the Oberth and Edelbaum maneuvers can produce a larger $v_\infty$ for a given $\Delta v$, as shown in Fig. 2. However, to obtain the benefit of reduced travel times, the destination distance must be much greater than the original orbital radius, since the Oberth and Edelbaum transfers require slow elliptical segments prior to the hyperbolic escape. A mission that calls for a high-speed escape from a lunar-like orbit around Earth[6] ($r_0 \approx 60$ Earth radii and $v_0 \approx 1$ km/s), or from a high orbit around any planet or asteroid, can benefit from employing the Oberth or Edelbaum maneuver.

Most mission concepts are constrained by the total $\Delta v$ available from the fuel on board. If employing either the Oberth or Edelbaum transfer, the inner radius $r_{\text{in}}$ of the intermediate orbit is limited by the radius of the central body (including its atmosphere), and in the case of the Sun, by heating and radiation considerations. (The *Parker Solar*

Table 2. Pros and cons of the 3-impulse (Edelbaum) escape.

| Pros | Cons |
|---|---|
| Can maximize $v_\infty$ for a given $\Delta v$ (and fuel expended). | Only advantageous for high orbits (Fig.1), fast escapes (Fig. 2), and distant destinations (Fig. 3). |
| In time reverse, can provide most efficient orbital capture. | |
| A plane change requires less fuel if executed at $r_{out}$ of intermediate orbit. | $r_{in}$ is limited by the central body's size and radiation environment. |
| Can be executed at any time (unlike gravity assist). | Requires two engine restarts. |
| Mission can be aborted back to the original circular orbit after the 1$^{st}$ or 2$^{nd}$ impulse, for a fraction of total $\Delta v_{Ed}$. | Spacecraft spends substantial time $\sim T_0$ in the vicinity of the original orbit before escaping. |

*Probe* makes perihelion passes as close as 10 solar radii ≈ 0.05 AU,[13] but an escaping spacecraft would only have to do so once.)

To date, no mission has used an Oberth or Edelbaum transfer to send a spacecraft to the outer solar system. This is mainly because the "Oberth advantage" over direct escape works best for values of total $\Delta v \approx v_0$ (≈ 30 km/s for a heliocentric orbit at $r_0$ = 1 AU), which is not currently attainable by chemical rockets. Some mission concepts have proposed hybrid gravity-assist/Oberth trajectories that use a carefully timed fly-by of Jupiter ($r_{out}$ = 5.2 AU) to provide some of the retrograde $\Delta v_{Ed2}$ necessary to cause the spacecraft to fall in close to the Sun.[14]

Thus far we have restricted our analysis to coplanar orbits and impulses. However, if one desires to change the plane of the escape hyperbola, the $\Delta v$ (and fuel cost) to do so is greatly reduced by applying the plane-change impulse simultaneously with the second, retrograde impulse of the Edelbaum transfer at $r_{out}$, where the spacecraft is moving slowest.

An important application of the 3-impulse Edelbaum maneuver is in time reverse, for efficient arrival and capture into a chosen circular orbit when a spacecraft approaches a planetary body at high relative speed. For approaches with $v_\infty > v_0$, the optimal strategy is shown in Fig 1(c) by reversing the direction of the arrows, and consists of a braking impulse at periapsis (which could be partially achieved by repeated aerobraking passes) into an eccentric orbit, followed by a periapsis-raising prograde impulse, and a third, "circularization burn".

For student discussions, Table II summarizes pros and cons of the Edelbaum transfer compared to direct escape and gravity-assist ("slingshot") trajectories.[15] Students can explore how the functions in Figs. 2 and 3 change as they vary $r_{in}$, $r_{out}$, and total $\Delta v$ to plan their own escape and approach maneuvers. They can also extend the analysis presented here to incorporate noncircular initial orbits, which can alter the choice of fuel-optimal transfer strategy. Trajectories can be simulated using software such as *Systems Tool Kit* (STK),[16] NASA's *General Mission Analysis Tool*,[17] *Orbiter*,[18] or *Kerbal Space Program*.[19] An STK animation of a race to escape between spacecraft employing the strategies discussed in this paper is provided online.[20]

## ACKNOWLEDGMENTS

Thanks to San Diego State University Library staff for assistance with locating aerospace journals published before 1970, to an anonymous referee for suggestions, and to Donna and Laura Edelbaum for encouraging this work.

## APPENDIX: RELATING POSITION AND TIME FOR A HYPERBOLIC TRAJECTORY

How does an orbiting object's position change with time? Many textbooks derive equations for this *Kepler problem* using a geometrical method similar to that first described by Johannes Kepler himself.[21] This requires the definition of new angular measures (the mean, true, and eccentric "anomalies") to locate the object on its conic section path. Such an approach makes sense for closed periodic orbits, and can be extended to unbound hyperbolic paths.[22] However, for our purposes we desire a direct relationship that gives the time taken for an escaping object to reach a given distance $r$ from the central body, starting from periapsis. Here we derive such a relationship using only conservation laws that will be familiar to students of classical mechanics.

### 1. The radial Kepler equation

We can rewrite the conserved specific orbital energy of Eq. (1) by separating the spacecraft's velocity into radial and azimuthal components $v_r$ and $v_\theta$, respectively,.

$$\varepsilon = \tfrac{1}{2}v_\infty^2 = \tfrac{1}{2}v_r^2 + \tfrac{1}{2}v_\theta^2 - \frac{GM}{r}. \quad (A1)$$

Use Eq. (3) to express $v_\theta$ in terms of the conserved specific angular momentum $h$ to form an expression for the radial speed as a function of distance,

$$v_r^2 = v_\infty^2 + \frac{2GM}{r} - \frac{h^2}{r^2}. \quad (A2)$$

The last "centrifugal barrier" term on the right-hand side reduces the radial speed as the spacecraft approaches the central body. At the periapsis distance $r = r_p$, we have $v_r = 0$, so from Eq. (A2) we can express the specific angular momentum $h$ in terms of $r_p$ and $v_\infty$, i.e.

$$h^2 = v_\infty^2 r_p^2 + 2GMr_p. \quad (A3)$$

Substitute for $h$ back into Eq. (A2) to obtain the differential equation

$$v_r = \frac{dr}{dt} = \sqrt{v_\infty^2 + \frac{2GM}{r} - \frac{v_\infty^2 r_p^2 + 2GMr_p}{r^2}}. \quad (A4)$$

While there is no closed-form solution for $r(t)$ when $v_\infty > 0$, this equation can be integrated numerically to give radial position versus time for an infalling or escaping object.[23]

Alternatively, to find the time since periapsis as a function of radial distance from the central body, we can integrate the reciprocal of Eq. (A4),

$$\frac{dt}{dr} = \frac{r}{\sqrt{v_\infty^2(r^2 - r_P^2) + 2GM(r - r_P)}}, \quad (A5)$$

with $t(r_p) \equiv 0$. The solution for $v_\infty > 0$ and $r \geq r_P$ can be written as

$$t(r) = \frac{GM}{v_\infty^3}\left[\sqrt{\left(1 + \frac{rv_\infty^2}{GM}\right)^2 - \left(1 + \frac{r_P v_\infty^2}{GM}\right)^2} - \cosh^{-1}\left(\frac{1 + \frac{rv_\infty^2}{GM}}{1 + \frac{r_P v_\infty^2}{GM}}\right)\right]. \quad (A6)$$

This radial version of the hyperbolic Kepler equation also encompasses purely radial trajectories with $h = 0$ and hence $r_p = 0$.[24] For comparison with traditional derivations,[22] the *hyperbolic anomaly* is given by the inverse hyperbolic cosine term in Eq. (A6).

### 2. Plotting the hyperbolic escape trajectory

For nonradial trajectories ($r_p > 0$), the spacecraft's angular position $\theta$ can be described in polar coordinates in terms of $r$ and the conic section's semi-major axis $a$ and eccentricity $e$,

$$r = \frac{a(1-e^2)}{1 + e\cos\theta} \text{ where } |\theta| < \cos^{-1}\frac{1}{e}, \quad (A7)$$

with $\theta = 0$ at the periapsis distance $r_P = a(1-e)$; the bounds on $\theta$ constrain the object's position to the correct branch of the hyperbola between its asymptotes. The geometrical parameters $a$ and $e$ are obtained from $v_\infty$ and $r_p$ using the standard results[22]

$$a = -\frac{GM}{2\varepsilon} = -\frac{GM}{v_\infty^2}, \text{ and } e = 1 - \frac{r_P}{a} = 1 + \frac{r_P v_\infty^2}{GM}, \quad (A8)$$

where for all hyperbolae $a < 0$ and $e > 1$. Examples of such paths are shown in Fig. 1

[a)] Electronic mail: philip.blanco@gcccd.edu
[b)] Electronic mail: mungan@usna.edu

1. I. Newton, *A Treatise of the System of the World*, (F. Fayram, London, 1728), pp. 5-7.
2. W. Hohmann, *The Attainability of Heavenly Bodies* (NASA Technical Translation F-44, Washington, 1960).
3. F. W. Gobetz and J. R. Doll, "A survey of impulsive trajectories," AIAA J. **7**, 801-834 (1969).
4. See P. Blanco, "A discrete, energetic approach to rocket propulsion," Phys. Educ. **54**, 065001 (2019). Here we define $v_{ex}$ such that the mechanical energy input per unit fuel mass expelled is given by $\frac{1}{2}v_{ex}^2$.
5. H. Oberth, *Ways to Spaceflight* (NASA Technical Translation F-622, Washington, 1972), pp. 194-217.
6. R. Heinlein, *The Rolling Stones* (Scribner, New York, 1952).
7. P. Blanco and C. Mungan, "Rocket propulsion, classical relativity, and the Oberth effect," Phys. Teach. **57**, 439-441 (2019).
8. T. Edelbaum, "Some extensions of the Hohmann transfer maneuver," American Rocket Society J. **29**, 864-865 (1959).
9. P. Blanco, "Slow down or speed up? Lowering periapsis versus escaping from a circular orbit," Phys. Teach. **55**, 38-40 (2017).
10. T. Edelbaum, "How many impulses?" Astronautics and Aeronautics **5**, 64-69 (1967).
11. L. Ting, "Optimum orbital transfer by several impulses," Astronautica Acta **6**, 256–265 (1960).
12. M. Pontani, "Simple method to determine globally optimal orbital transfers," J. Guidance Control & Dynamics **32**, 899-914 (2009).
13. D. Longcope, "Using Kepler's laws and Rutherford scattering to chart the seven gravity assists in the epic sunward journey of the Parker Solar Probe," Am. J. Phys. **88**, 11-19 (2020).
14. A. Hibberd, A. Hein, and T. Eubanks, "Project Lyra: catching 1I/'Oumuamua – mission opportunities after 2024," Acta Astronautica, **170**, 136-144 (2020).
15. J. A. van Allen, "Gravitational assist in celestial mechanics - a tutorial," Am. J. Phys. **71**, 448–451 (2003).
16. To obtain free educational licenses for STK, visit http://www.agi.com/education.
17. http://opensource.gsfc.nasa.gov/projects/GMAT.
18. Online at http://orbit.medphys.ucl.ac.uk.
19. Online at http://www.kerbalspaceprogram.com.
20. See the supplementary materials online.
21. F. G. Orlando, C. Farina, C. A. D. Zarro, and P. Terra, "Kepler's equation and some of its pearls," Am. J. Phys. **86**, 849-858 (2018).
22. C. A. Kluever, *Spaceflight Dynamics*, 2nd ed. (Wiley, Hoboken, 2018).
23. P. Blanco, "Angular momentum and Philae's descent," Phys. Teach. **53**, 516 (2015).
24. C. E. Mungan, "Radial descent of an energetically unbound spacecraft towards a comet," Phys. Teach. **53**, 452-453 (2015).